Femtosecond laser micromachining of diamond: current research status, applications and challenges

Bakhtiar Ali Khan [a, b], Igor V. Litvinyuk [b, c], Maksym Rybachuk [a, b] *

[a]  School of Engineering and Built Environment, Griffith University, 170 Kessels Rd, Nathan QLD 4111, AUSTRALIA

[b]  Centre for Quantum Dynamics and Australian Attosecond Science Facility, Griffith University, Science Road, Nathan QLD 4111, AUSTRALIA

[c]  School of Environment and Science, Griffith University, Nathan QLD 4111, AUSTRALIA

* Corresponding author. Tel: +61 (0)7 5552 7268. E-mail: *m.rybachuk@griffith.edu.au*

TITLE

Femtosecond laser micromachining of diamond: current research status, applications and challenges

ABSTRACT

Ultra-fast femtosecond *(fs)* lasers provide a unique technological opportunity to precisely and efficiently micromachine materials with minimal thermal damage owing to the reduced heat transfer into the bulk of the work material offered by short pulse duration, high laser intensity and focused optical energy delivered on a timescale shorter than the rate of thermal diffusion into the surrounding area of a beam foci. There is an increasing demand to further develop the *fs* machining technology to improve the machining quality, minimize the total machining time and increase the flexibility of machining complex patterns on diamond. This article offers an overview of recent research findings on the application of *fs* laser technology to micromachine diamond. The laser technology to precisely micromachine diamond is discussed and detailed, with a focus on the use of *fs* laser irradiation systems and their characteristics, laser interaction with various types of diamonds, processing and the subsequent post-processing of the irradiated samples and, appropriate sample characterisation methods. Finally, the current and emerging application areas are discussed, and the challenges and the future research prospects in the *fs* laser micromachining field are also identified.

CONTENTS                                                                                                                pages

# 1    INTRODUCTION

Diamond remains a material of choice for applications where extremes of thermal conductivity (2000 W m$^{-1}$ K$^{-1}$), Young's modulus (1220 GPa), mechanical surface and bulk hardness (100 GPa), optical index of refraction from ultra-violet (UV) to infra-red (IR) range (2.41), dielectric constant (5.7) and electrical resistivity ($10^{13}$-$10^{16}$ Ohm-cm) and low thermal expansion (1.1 × $10^{-6}$ K$^{-1}$) are required [1-4], such as in ultra-high precision machining tools [5], micro-electro mechanical system (MEMS) and MEMS components [6], robust optical gratings [7], inert and environmentally stable medical coatings [8-10], and electronic devices for harsh environments [11]. The availability of synthetic diamonds in different grades (*i.e.,* 'a' and 'b' grades), in various degrees of crystallinity, purity, degrees and types of doping, physical shapes and forms and, subsequently, at highly varied prices, made synthetic diamonds both more attractive and more economically accessible for a wide range of industrial applications compared to their natural counterpart [12, 13]. This also includes the engineering applications where diamond and its derivatives are used as machining tools [14]. The advantage of using tailor-made diamond products lies in its exceptionally high mechanical hardness that originates from directional *sp$^3$*- hybridised bonds arranged in tetrahedral configuration that compose a single crystal [15]. It is the world's most desirable, albeit most notoriously known as 'hardest-to-machine' material, since there are no conventional material removal processes (MRPs) that are capable to effectively shape diamond crystals with an acceptable precision and accuracy at the nano- and/or at micro- scale. Historically, a number of MRPs have been employed to process diamond into the desired shapes including electrical discharge [16, 17] and abrasive waterjet machining [18], mechanical grinding [19], and laser machining [20, 21]. These methods utilise the ability of diamond's intrinsic *sp$^3$*- hybridised bonds to become converted, through externally applied energetic processes, into much weaker, *sp$^2$*-hybridised bonds (as in graphite) and after their subsequent removal, impart a final shape to the product [15]. Among these methods, laser machining (aka. laser processing) has been widely and, so far, the most successfully used MRP [21-23], since an appropriate selection of wavelengths, pulse durations and power results in a high quality, tailored materials' surface and precise volumetric processing down to a few nanometres precision [24, 25].

Ultra-fast laser micromachining using femto-second (*fs*) pulses offers many distinct advantages compared to any other laser machining processes. For diamond, the advantages are the following: firstly, the interaction time of laser light with diamond sample is very short (*e.g.,* pulse duration is less than a picoseconds) resulting in a low volume, insignificant amount of heat dissipating into the surrounding bulk of the matter from the irradiated area, which in turn minimises the thermal and mechanical stresses within the surrounding area, the latter can also be further minimised by reducing the laser beam spot size. Secondly, *fs*-laser light interactions with diamond crystal are independent of its orientation and, subsequently, no specific requirements for diamond sample positioning (in a sample holder) prior to laser processing are required. Thirdly, the laser beam used in material processing, as a 'machining tool' depends only on the optical quality of the beam and therefore, it is not affected by the sample-to-tool interaction, unlike in other conventional MRPs. Owing to these characteristics *fs* laser micromachining is able to successfully support several high impact engineering applications that exploit devices' complex surface topography and miniaturised features, including electron emitting devices for high-power/high-speed electronics [26, 27] and robust microfluidic devices [28] made entirely of diamond, fabrication of which would not be technologically attainable using conventional machining methods.

This article reviews the state of knowledge and the research work performed to date on *fs*-laser micromachining of diamond and addresses several important questions:

1) What light-induced changes occur in diamond and how these changes affect diamond properties and its surface topography (Section 2)?

2) Why *fs*-laser machining should be preferred to precisely micromachine diamond, despite other (nano- and pico- second) laser processing methods currently available (Section 3)?

3) What are the processing parameters to machine diamond using a laser light, how to select and tailor the laser system and sample processing parameters, and how to select the most appropriate analytical methods to study the topography and chemical structure of machined diamond crystals (Section 4)?

4) What are the current and forthcoming future applications for machined diamond, and what are the future research prospects in this developing field (Section 5 and 6)?

We expect this brief review to appeal to the entire carbon research community and the broader laser processing/laser machining community as a whole. Given that the article addresses an intricate combination of diamond physicochemical properties, laser system design, photo-induced ablation mechanism of carbon materials, it will also appeal to laser engineers, optical physicists and materials engineers in other fields.

## 2  LIGHT-INDUCED STRUCTURAL CHANGES IN DIAMOND

### 2.1  Diamond absorption of laser energy

Laser ablation (aka. photoablation) is the process of removing material from a solid or, occasionally, liquid surface by irradiating it with a laser beam. At low laser flux, the material is heated by the absorbed laser energy and evaporates (or sublimates). The nature of the effects of the interaction of the laser light with a solid depends on many factors, among which the most relevant are the laser pulse duration ($\tau$), wavelength ($\lambda$), laser power ($P_{avg}$) (*i.e.,* beam energy density), exposure time and repetition rate ($R_p$) (in case of pulsed lasers), the beam characteristics, the optical characteristics of the applicator and physical properties of a solid. Inside the solid, light can be reflected, transmitted, scattered and absorbed. Only absorbed energy can produce the desired ablation effects in a solid, whereas reflected, transmitted and scattered light could affect the shape, the extension and the position of the warmed-up volume within the ablated area. The depth to which a certain amount of energy can ablate a solid depends on its absorption coefficient (*i.e.,* attenuation coefficient) and its enthalpy of vaporization (*i.e.,* heat of vaporization), which for diamond is around 711 kJ/mol [29]. The phase diagram of carbon, given in Fig. 1, shows that at ambient pressure diamond when subjected to heat first, converts to graphite and, eventually sublimates bypassing the liquid state [30].

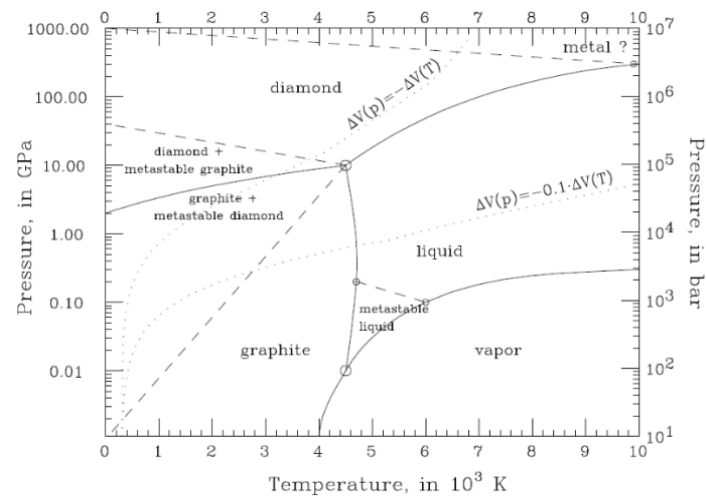

***Fig. 1.*** *Phase diagram of carbon [30].*

The ablation begins above the threshold fluence (TF), which predominantly depends on material properties and the presence of intrinsic and extrinsic defects in the bulk and at/on the surface of an ablated solid. The ablation TF also varies with the pulse duration (see Fig. 2). Ablation threshold for the synthetic diamond in the ultrashort *fs* regime is about 3 J/cm$^2$ [31]. Laser-based MRPs often employ the application of multiple pulses to reduce the ablation threshold at the ablated surface area owing to the defect accumulation at the uppermost surface of the ablated area [29]. Notably, the material removal rate (MRR) per pulse above the ablation threshold, in effect, follows the Beer–Lambert law, with optical attenuation of the ablating laser light directly related to the optical path length and absorptivity of the ablated specimen [32].

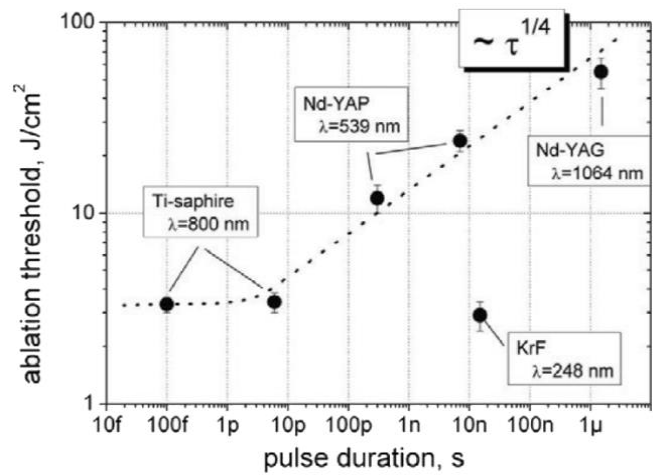

*Fig. 2. Ablation threshold fluence vs. pulse duration [31].*

There is a considerable difference between the optical absorption of a short-pulse, such as a nano-second and a pico-second (*i.e.,* ns and ps, respectively) and an *fs*-pulsed laser radiation in diamond [29]. The laser energy absorption in pure, single crystal diamond (SCD) occurs through the electronic excitation across the σ - σ* bond forming an exceptionally wide bandgap, $E_g$, which depending on crystallinity, varies between 5.0 and 5.5 eV [15]. The $E_g$ can only be bridged by high energy UV photons ranging from approximately 250 nm (for a 5.0 eV $E_g$) down to ~ 225 nm (for 5.5 eV) diamond crystals. However, conventional laser ablation systems normally use a visible 532 nm (green) or a near infra-red (*N-IR*) 1064 nm light for laser machining processes and, as a result, are unable to sublime pure diamond efficiently.

Ablation in diamond is strongly affected by the presence of intrinsic and extrinsic defects including ad-atoms, interstitial or substitutional defects in the crystal lattice, point, line (*e.g.,* grain boundary), volume and surface defects [15]. These defects collectively reduce the TF and high defect concentrations are associated with low ablation TFs in diamond, which is advantageous for laser machining. In an instance when defects/impurities are absent altogether, multiple photons are able to excite an electron in a process known as multiphoton absorption resulting in improved ablation rates [15], the latter, requires a high spatial and temporal photon density of an *fs*-pulsed laser with a focused beam [29]. Ramanathan *et al.* [33] reported that even though the optical transparency of diamond is quite high for a vis – N-IR laser light, high peak power enables a stable absorption through a multiphoton ionization and absorption at defects in diamond. Kononenko *et al.* [34] demonstrated that a precise multiphoton absorption prevails in laser fluences over a wide range. Both studies show that a severe non-linear transformation is experienced by *fs*-pulsed laser beam at path lengths of tens of microns or longer, which determines the energy absorbed by the diamond's crystal lattice.

In the process of ablation the $sp^3$- tetrahedral phase of diamond is generally removed (*i.e.,* sublimed) through a combined process of graphitization, in which the $sp^3$ phase is thermally converted into olefinic and/or aromatic $sp^2$ fraction , and ablation and oxidization processes [35]. The heat affected zone (HAZ) can be effectively minimised by using *fs*-pulsed laser irradiation owing to its ultra-short interaction time, which promotes the formation of a narrow ablation profile with a minimal pile up and skirting features [29, 33]. Ogawa *et al.* [36] noted that *fs*-pulse transfers its energy via a photon-electron coupling fully prior to phonon-driven diffusion, which induces an ablation process that is, unique for *fs*-light, occurs without a significant graphitisation. It has been proposed that diamond sublimation occurs when the peak temperature reaches above 4800 K, whereas it graphitizes when the temperature in HAZ remains between 2000 K and 4800 K [15].

Additionally, it is now known that the MRR increases with an increase of incident *fs*-laser ablation energy, however only a minimal change in the thickness of an $sp^2$-graphitised layer is observed [36]. Notably, all types of laser irradiation reduce optical transmittance of diamond's surface owing to the optical breakdown of the bulk transparent solid, the latter, in fact further reduces the suitability of laser ablated diamonds for optical applications [37]. Nonetheless, the use of *fs*-pulsed light significantly minimises the severity of the observed optical breakdown as ultra-short pulse durations in *fs*-generated pulses are several orders of magnitude lower in frequency compared to the electron-lattice relaxation time ($10^{-10}$-$10^{-12}$ sec) [38].

## 2.2  Photoinduced changes in diamond properties

In diamond, carbon, C, atoms form an *sp³* hybridized directional covalent bonds with the neighbouring atoms in tetrahedral orientation with an *s* orbital overlapping three *p* orbitals and, hereby, providing strength to the *σ* bonds at an enclosed C–C–C angle of 109.47 ° and C–C bond length of 1.54 Å and a bandgap of 5.5 eV [39]. Laser irradiation cleaves one (or at higher fluences, a few) of covalent C–C *σ* bonds allowing for three covalently-bonded electrons to re-arrange themselves to a planar orientation by slightly opening the enclosed C–C–C bond angle to 120 ° and reducing the C–C bond length to 1.42 Å. The fourth electron becomes delocalised over the entire *π* - *π* sheet, overall forming an *sp²* hybridized bonding arrangement as in graphite. This *sp³* to *sp²* phase transition drastically reduces the bandgap to -0.04 eV. Notably, the *sp³* to an *sp²* phase transition performed in a controlled manner opens a unique opportunity for the development of 'all-carbon' electronic, optical and even quantum devices [40, 41] owing to the extreme differences in atomic structure, electron arrangement and chemical configuration and interaction mechanism between the two most common carbon allotropes.

In the *fs* regime, when diamond is irradiated with an applied fluence above its ablation threshold of approximately 3 J/cm², ablation with minimal graphitization is observed (at depths fewer than 50 nm) [36], whereas at fluences below the ablation threshold (*i.e.,* < 3 J/cm²) no ablation is occuring and mostly graphitization is obserbved marked with a formation of an *sp²* phase. This phenomenon is graphically shown in Fig. 3a, where the optical transmission of chemical vapor deposited (CVD) diamond rapidly falls at 0.3 J/cm² (well below diamond's ablation threshhold) indicating the start of graphitization process [12]. It has been suggested, that the bulk transformation process from an *sp³* to an *sp²* bonding proceeds with the direct photodamage of the diamonds' *sp³* lattice creating an *sp²* nuclei and subsequent heating of newly formed and embedded *sp²* graphitic centres in the irradiated *sp³* volume [42]. Fig. 3b shows the reduction in total optical transmittance attributed to an increasing local graphitization relative to the projected optical power (the number of laser shots at different incident fluences) for a *ps*-laser light [43]. Notably, there is an initial photonic accumulation period observed, where the energy density remains low for relatively low laser fluences.

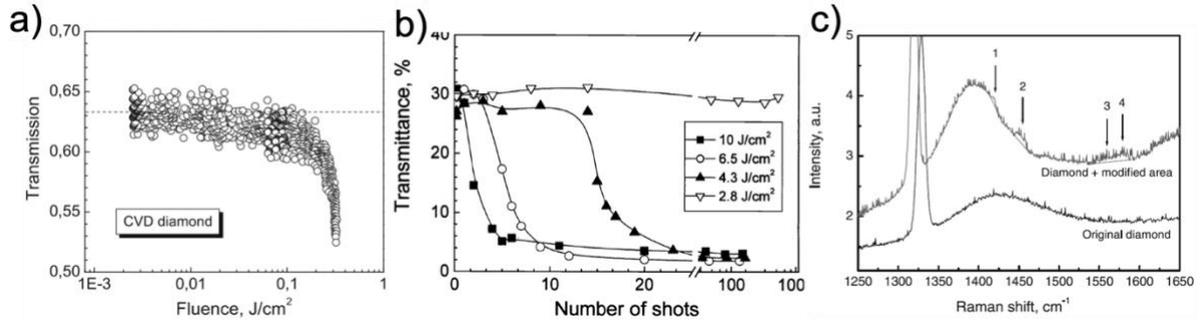

***Fig. 3.*** *a) Optical transmission of CVD diamond specimens vs. incident laser fluence (120 fs, 800 nm) [12], b) changes in the optical transmittance of CVD diamond vs. number of laser shots at different fluences (220 ps, 539 nm) [43], c) Raman spectra of original CVD diamond (bottom) and laser modified region (top) [12].*

The structure of downconverted $sp^2$ phase that is formed following the laser irradiation of diamond samples in air is never purely crystalline but is highly amorphous. For example, a 532 nm Raman spectra of the original nitrogen, N, doped CVD diamond commonly displays the characteristic $sp^3$ vibrational mode at ~1331 cm$^{-1}$ and, a low intensity broad mode at ~1420 – 1425 cm$^{-1}$, which is commonly atributed to an N-doped CVD, as shown in Fig. 5 [12]. Laser irradiation gives rise to additonal vibrational modes attributed to microcrystalline graphite at ~1417 cm$^{-1}$ (peak 1, Fig. 3c), amorphous $sp^2$ carbon at ~1450 cm$^{-1}$ (peak 2, Fig. 3c) and distorted microcrystalline graphite at ~1560 cm$^{-1}$ (peak 3, Fig. 3c) and 1580 cm$^{-1}$ (peak 4, Fig. 3c) [44, 45]. The $sp^2$ rich phase formed in diamond following laser irradiation is mostly composed of a mixture of an $sp^2$ aromatic clusters and $sp^2$ olefinic chains [46, 47] and often contains poly(*p*-phenylene vinylene) and poly(*p*-phenylene) $sp^2$ fractions, analogous to those found in amorphous carbon materials synthesised in hydrocarbon plasmas [48-50].

Following the preliminary $sp^3$ to $sp^2$ phase conversion and the formation of a graphitised nuclei the photoablation of diamond proceeds as a combined process of vaporization and oxidation. The vaporization process occurs both in air and in vacuum, whereas the oxidation process occurs only in the presence of oxygen. Komlenok *et al.* [51] reported that at high laser fluences of 1.2 - 3.8 J/cm$^2$ that are a fraction below and slightly above its TF, diamond can be sublimed at a rate of 30 - 200 nm per pulse in both air and vaccum owing to vaporization driven photoablation. At low fluences of 0.34 - 0.55 J/cm$^2$, which are an order of magnitude lower than the diamond's TF, the siblimation is minimal at 0.01 nm per pulse as the process is mainly driven by oxidation [51] and, no submimation occured at low fluences in the absence of oxygen (in vaccum).

The photoelectric properties of diamond, including optical absorption and photon-to-electron quantum efficiency, depend strongly on its intrinsic crystalline structure and the presence of impurities. Owing to its high refractive index, $n$, that is 2.4 at 590 nm [52], photocurrent in a pure diamond crystal can be induced by photon absorption in the UV range (below 225 nm) [53], however in an instance when impurities and defects are present, - in the visible range (415-478 nm) [54]. This is a very important experimental observation indicating that the presence of defects in diamond increases the photoconductivity by enabling light of lower energy to be absorbed in the material. Brecher *et al.* [15] recently reported that UV induced photoconductivity over a prolonged period reduced the sensitivity of diamond-based UV detectors. The sensitivity could be fully restored by thermal annealing of diamond crystals in white light [15].

## 2.3    Photoinduced changes in diamond surface topography

Descriptions of surface topography in applications to laser-processed materials normally refer to the shape of the surface profile (*i.e.,* waviness, type of surface finish) and an average arithmetic surface roughness, $R_a$, value, which are known to significantly impact the material's bulk properties and its suitability for applied applications. Irradiation of materials surface changes the surface topography depending on the shape, intensity and polarization of an incident laser beam. Laser irradiation results in the formation and growth of an $sp^2$ graphitized region immediately within the HAZ in diamonds at lower fluence, which can be rapidly and easily sublimated at higher fluence [55], depending on the pulse width of the incident laser pulse. The use of *fs*-pulse light is associated with the reduced heat transfer into the bulk of the work material [56], sharp thermal profiles and high precision machining [33, 55] compared to conventional *ns*- and or *ps*-laser processing methods. Using an *fs* beam with a circular polarization enables effective micromachining of various spherical, rounded, globular and conical features on the surface of the ablated solid, while preserving the $sp^3$ phase composition, including curved profiles when accelerating beams are employed [57]. The use of laser beam with a linear polarization generates graphitic periodic surface structures over the irradiated surface [58, 59], including ripple-like features with a characteristic periodicity, close to the wavelength of the incident light. The latter phenomenon arises from the interference of the incident laser beam with the scattered waves from the surface of an ablated solid. The interference between the incident light wave and the surface plasmon excitation is responsible for the generation of the surface structures that display periodicity patterns close to or, sometimes, less than the laser incident wavelength. The latter is dependent on the intrinsic $R_a$ values and surface features of the irradiated workpiece [59]. It is important to point out that the periodic ripples of the surface of an ablated solid display the periodic features that are much smaller compared to the irradiated laser wavelength range, are smaller for *fs*-laser irradiated solids compared to *ps-* or *ns-* irradiated materials [44, 45, 60, 61]. Fig. 4 shows surface morphology of CVD diamond samples irradiated with linearly polarized *fs- l*aser beam at fluences near the ablation threshold (*i.e.,* 1.9 J/cm$^2$) [62]. The scanning electron microscopy (SEM) images of nano-sized gratings with regular ~170 nm periodicities are shown displaying abrupt verges between the grooves and ridges [62]. However, the use of a slightly higher fluence (2.8 J/cm$^2$) results in an increased periodicity of ~190 nm and a smoother profile of the gratings due to stronger thermal effect [62].

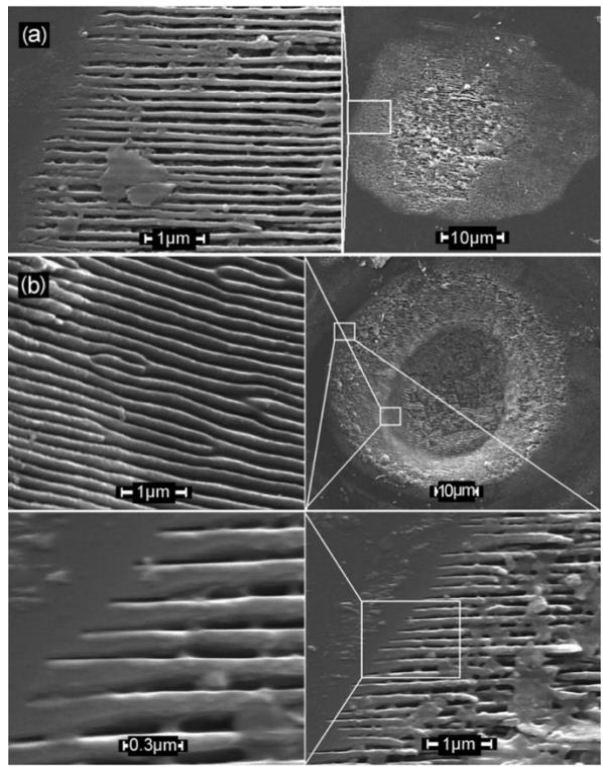

***Fig. 4.** SEM images of diamond irradiated by 800 nm fs laser light: (a) 3000 pulses at a fluence of 1.9 J /cm$^2$ result in a periodicity of ~170 nm, (b) 8000 pulses at a fluence of 2.8 J /cm$^2$ produce smoother periodicities of ~190 nm [62].*

Various features can also be laser written (aka. scribed) on diamond with a suitable periodicity and spacing for applied waveguide applications. Recently, Bharadwaj et al. [63] reported on successful fabrication of mid-IR waveguides by scribing two 40 µm waveguide lines on SCD, which were designed to operate at 2.4 µm and 8.6 µm wavelengths; the waveguiding was possible owing to an increased polarizability and a greatly reduced refractive index in the waveguide's $sp^2$ rich cladding sheath [64]. Laser irradiation can be tuned to reduce $R_a$ value of the machined workpiece, which is generally considered as an indicator of improved machining surface quality. Moreover, for applied applications it is important to maximise MRRs in laser ablation processes as a measure of improved machining efficiency. Yao et al. [65] have shown that in an instance, when nano-crystalline CVD diamond with is subjected to fs laser ablation, the $R_a$ and MRR are almost linearly co-dependent (Fig. 5; notably, the $R_a$ of as-grown CVD was 0.342 µm).

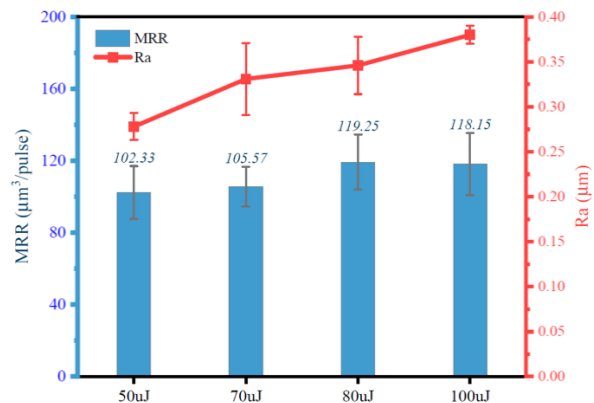

*Fig. 5. The $R_a$ and MRR vs. single pulse energies (100 fs, 800 nm, 1 kHz, $S_s$: 1 mm/s) [65].*

Fs laser irradiation enables precise modification of diamond surface profile depending on laser beam characteristics. However, the increase of MRR in laser machining adversely affects the surface roughness of processed samples, the latter effect to be minimized through a careful selection of process parameters and the use of appropriate to the sample ablation threshold parameters.

## 3 LASER MICROMACHINING OF DIAMOND

### 3.1 Conventional (nano-, pico- second) laser micromachining of diamond

Laser irradiation, a versatile tool for material processing, enables precise control over the amount of energy focused at a specific location on the sample by selecting appropriate to the sample process parameters [66]. Laser ablation processes normally include laser grooving, scribing, drilling and marking [10, 67], all of which are used to remove bulk of the material. The *ns-* laser ablation process is governed by electrons in an ablated media absorbing laser photon energy and reaching a thermal equilibrium with the lattice during the laser pulse duration [68]. Owing to a relatively short electron-lattice relaxation time of $10^{-10}$-$10^{-12}$ s, that is up to three orders of magnitude smaller, compared to *ns-* laser pulse duration [38], the *ns-* laser ablation processes are thermally destructive to the sample media as extended HAZs are generated in the process. Takayama *et al.* [69] classified four different forms of damage and their causes that occur during *ns-* laser processing of diamond (employing 15.6 ns, 532 nm, $P_{avg}$ > 1W system as a study model), namely: cracking, groove shape deformation, ripple formation, and debris deposition. Cracking resulted from a rapid temperature change, groove shape deformation resulted from an improved absorption of the plasma generated through *ns-* laser irradiation, surface ripples were formed by the *ns-* laser beam interference with the groove walls, whereas laser induced debris deposition was attributed to the use of sample-specific ablation regimes [69]. Cadot *et al.* [10] (see Fig. 6) have shown that graphitisation and $sp^2$ amorphization of diamond occurs at relatively low fluences for *ns*-pulsed laser as evidenced by appearance of characteristic disordered graphite D and G Raman modes at ~1350 cm$^{-1}$ and 1580 cm$^{-1}$, respectively [48, 49, 70], which are becoming broader and display higher intensities at higher fluences.

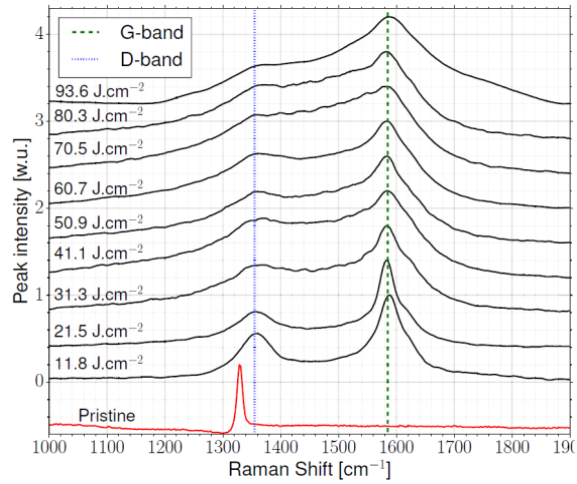

*Fig. 6.* *1064 nm Raman spectra of diamond irradiated with 30 ns pulsed laser at 35 kHz at various fluences (adapted from [10]).*

Kononenko *et al.* [71] reported that reduction of *ns*- pulse width resulted in less graphitization and the disordered graphitized layer thickness was found to be independent of the laser energy for fluence ranges above the ablation threshold, namely at or above $0.2\ \text{J/cm}^2$. Ohfuji *et al.* [21] found that the ambient conditions during *ns*-pulsed laser machining of PCD samples had a negligible effect on both the thickness of the final graphitized layer and the recorded MRR values. These works additionally confirm that the predominant mechanism of material removal in diamond is by graphitization and subsequent sublimation of the graphitized layer from the ablated volume/region, as noted ealier.

It has been reported that excimer lasers commonly operating in the range of 5 – 20 ns pulse lengths are capable of producing deep holes in diamond during laser drilling operations [72], the actual depth of the hole that can be created is, however, limited by the generated hydrocarbon plasma plume, which obstructs the laser energy from reaching towards the bottom of the hole and thus reducing the pressure available to expel the melted and graphitised material out. Typical ablation plasma attributes include characteristics such as ion density, ion flow velocity, free electron velocity as well as electron temperature [73]. Naturally, longer pulses generate plasma plumes of higher ion density and electron temperature and, therefore short pulse widths have been recommended for laser drilling processes where an increase in the depth of hole was required. Earlier works of Kononenko *et al.* [74] have shown that *ps*- pulsed laser drilling of CVD diamond generated less plasma plume while creating a deeper hole (*i.e.,* 500 μm) compared to a *ns-* pulsed laser used under similar conditions, which produced a much shallower hole (*i.e.,* 50 μm).

Additionally, it was also observed that a significantly higher threshold fluence (~17 J/cm$^2$) was required for the through-hole laser drilling process compared to the blind-hole shallow drilling process (2 J/cm$^2$) for the same workpiece material type. The latter observation was attributed to an increased photon absorption of dense plasma restricted by the wall of a deep hole. To curtail the photon scavenging effects of plasma plume during laser drilling of CVD diamond Migulin *et al.* [75] employed an additional oxidation potential of an $O_2$ jet to facilitate the removal of sublimed graphitised fractions from the ablated volume through CO- and $CO_2$ formation, and found the solution reasonably effective.

It is often noted that the thickness of the graphitized layer formed on diamond sample is inversely proportional to the projected *ns*-laser fluence since complete down-conversion from an *sp$^3$* to aromatic *sp$^2$* fraction is only possible when diamond sublimes above its ablation threshold. Mouhamadali *et al.* [76] showed that CVD diamond transformed to a 'thicker' amorphous *sp$^2$* graphite-like (*a*-C) when irradiated with a green (532 nm) 40 ns light at a relatively low fluence of 4.9 J/cm$^2$, whereas only a 'thin' *sp$^2$* aromatic phase was observed at much higher 15 J/cm$^2$ fluences. Additionally, multiple micro cracks on the surface of aromatic *sp$^2$* sites ablated at 15 J/cm$^2$ were observed, while only minor cracks were detected on *a*-C sites ablated at 4.9 J/cm$^2$ and the thickness of *a*-C layer was found to be ~1.5 thicker compared to *sp$^2$* aromatic phase.

*Ps*-laser machining was found to be more effective compared to *ns*-pulsed laser processes given a significantly shortened pulse durations and, subsequently, the likelihood of HAZ formation in the ablated volume. Such as Takayama *et al.* [77] performed a *ps*-pulsed laser irradiation (1030 nm, 800 ps) using variable repetition rates on a SCD sample to make a tool with a micro-grooved edge and found that 15.3 J/cm$^2$ fluence at 100 kHz afforded rapid machining with limited graphite deposition and devoid of any edge cracking. Guo *et al.* [78] investigated the effect of *ps*-laser-induced micro-grooves on the grinding performance of coarse-grained diamond wheels for optical glass surface grinding and reported that the laser micro-structuring reduces the subsurface damage depth, *d'*, effectually from 5 to 1.5 □m. Both the *d'* and an $R_a$ values were found to be lower with reducing the micro-groove spacing. Zhang *et al.* [79] produced macro-patterns on the diamond grinding wheel surface to improve the grinding performance and reported that the grinding temperature and sub-surface damage decreased significantly for macro-structured grinding condition. The normal and tangential grinding forces were reduced by approximately 15% for *ps*-laser macro-structured grinding wheels compared to the conventional grinding wheels.

Various methods have been proposed to improve the laser machining performance. For instance, Park *et al.* [80] performed micromachining with a *ns*-pulsed excimer laser and an assisted compressed auxiliary gas in the machining of diamond films, and observed that the laser beam breaks the atomic bonds in the work material and the resultant plasma plume gets expelled away by an assisted gas jet, improving machining quality of the diamond surface. Additionally, both the *ns*- and *ps*-laser processing has been successfully combined with precision grinding to improve the machining performance by several researchers, such as Brecher *et al.* [15, 81] employed a *ns*-laser combined with grinding for fabrication of PCD turning tools. In this process, the surface was micromachined using *ns*-laser irradiation to remove the bulk of the material, assisted by conventional grinding process to remove the residual. The cutting forces, the magnitude of grinding wheel vibration and the overall processing time were significantly reduced by approximately 50 % compared to traditional grinding. Likewise, Yang *et al.* [82] performed the laser induced graphitization of CVD diamond into predominantly aromatic graphite, which was subsequently removed via precision grinding process.

A fundamental problem with a *ns*-laser ablation lies in its inability to generate suitably high ablation rates and acceptable machined surface quality, the problem is also extended to the application of a *ps*-laser, albeit to a lesser extent. The cracking tendency, the generation of pervasive HAZs and poorly controlled graphitization of machined diamond surfaces are mostly attributed to relatively long (i.e., *ns*- and *ps*-) pulse durations compared to electron-lattice relaxation times of $10^{-10}$-$10^{-12}$ s [83]. In particular, *ns*- process generates more recast layers, debris and HAZ compared to the *ps*-laser process owing to the extended photon-lattice interaction time [84] and is associated with significant and stable plasma formation at the surface of the ablated media [85]. It is hardly a materials processing method of choice in instances where integrity, composition and properties of tetrahedral *$sp^3$* organisation in diamond are to be preserved during the post laser machining.

## 3.2  *Fs*- laser micromachining of diamond

The first reported attempts to develop an ultrashort-pulsed laser with a pulse duration of less than 0.1 ps were made in early 1980s by Fork *et al.* [86] and it has taken over two decades to recognise an opportunity offered by ultrashort laser technology for precise photoablation of diamond. In *fs*- laser machining, optical energy is transferred to the target media through photoinduced optical breakdown, in which a majority of electrons are ionized resulting in a structural and, in the instance of diamond ablation, a phase (*i.e.*, $sp^3$ to $sp^2$) transformation and, subsequent ablation of the irradiated volume. In a highly transparent material, such as dopant-free pure diamond, the energy absorption must be non-linear due to the insufficient incident photon energy avaiable for ionisation [87, 88], and for a non-linear absorption to occur, the strength of the electric field in the laser pulse needs to be equal to the electric-field strength, something in the order of $10^9$ V m$^{-1}$ (approximately corresponding to laser intensity of $5 \times 10^{20}$ W m$^{-2}$) [89]. High intensity and tight focusing are required to achieve electric-field strengths of such a magnitude. The non-linear absorption and tight focusing confine the absorption inside the bulk of the material to the focal volume preventing the absorption at the surface and thus yielding micromachined volumes as small as 0.008 μm$^3$ [90].

The ultrashort laser pulses deposit the laser energy in thin layers with the thickness of $1/\alpha$, where $\alpha$ is the optical absorption coefficient and, sublime the material through a direct vaporisation from the surface [91], which for diamond occurs through conversion of $sp^3$ into a full- or a partial- $sp^2$ phase configuration and, subsequent sublimation of $sp^2$ phase from the irradiated volume. The sublimation occurs only at or above the sample ablation threshold, and it is the critical threshold fluence, which causes the material removal within the irradiated spot area and the focal volume in the sample. The fluence, however, should not exceed a certain value, specific for a given diamond sample, to avoid the thermal damage and an uncontrolled increase of the HAZ in the focal volume. The *fs*-pulsed lasers are able to generate extremely high-power density (up to a few GW) at relatively low average laser power (as low as 100 mW) [92], high enough to dissociate C–C covalent bonds in a diamond lattice. Graphitization occurs by electrons absorbing energy through an inverse bremsstrahlung process and, thus facilitating a transition from an $sp^3$ tetrahedral to an $sp^2$ aromatic and/or an $sp^2$ olefinic bonding state [93]. The $sp^3$ to $sp^2$ transition increases C–C interatomic distances, lowers the density of states (DOS) and changes the physico-chemical properties of the irradiated solid [47-50]. Notably, the *fs*-lasers operating at short pulse durations (*e.g.*, in $10^{-15}$ s range) enable precise manipulation of DOS owing to a relatively long electron-lattice relaxation times of $10^{-10}$-$10^{-12}$ s, reduce the probability of HAZ formation, sublime diamond with minimal thermal damage and enable the manufacture of surface structures with clear and well-defined edges as shown in the Fig. 7(a) and 7(b) and only displaying nanoscale surface ripples Fig. 7(c).

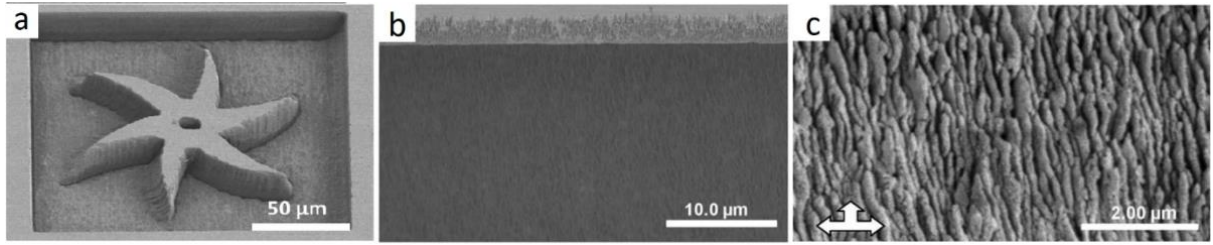

*Fig. 7.* SEM images of the high pressure high temperature (HPHT) synthetic SCD surface with a pulse train of 200 fs- laser pulses at a rate of 250 kHz: (a) curved structures machined at a laser pulse energy of 1.2 mJ, (b) image of the machined surface at pulse energy of 840 nJ, (c) magnified image of (b) (adapted from [94]).

*Fs*-laser micromachining involves several physical processes with fairly well-defined timescales as shown in Fig. 8(a) [95] and 8(b) [89].

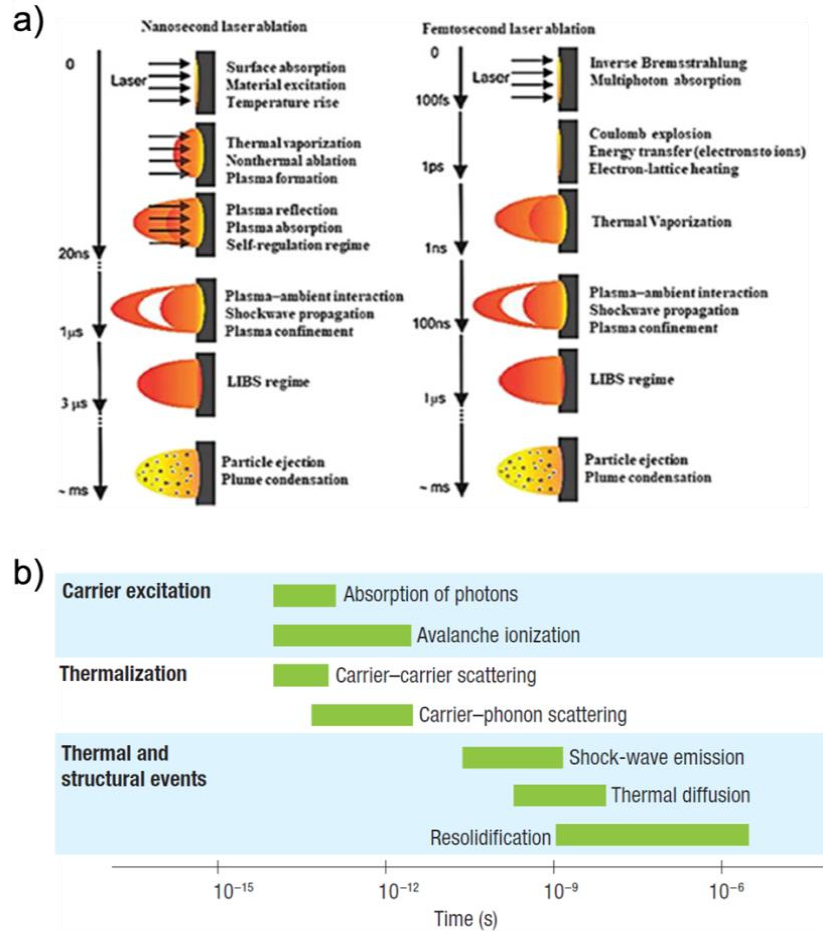

***Fig. 8.*** *a) Ablation mechanism of a short pulse (i.e., ns- laser ablation), adapted from [95] and b) timescale of physical phenomena associated with the interaction of an ultra-short fs-pulse laser ablation, adapted from [89].*

Namely, it takes a few ps for electrons to transfer absorbed optical energy to the lattice, it takes few ns for a dynamic shock/pressure wave to separate hot and dense focal volume [96, 97] and, an additional few µs for the thermal energy to diffuse out of the focal volume. These processes result in a non-thermal ionic motion at a sufficiently high energy to leave behind permanent and lasting structural changes [98]. There is a fundamental difference between the *fs*-photon induced damage and that of the *ps*- or *ns*- ones [89]. For sub-*ps* pulsed lasers, the timescale over which an *fs*-pulse excites an electron is much smaller than the electron–phonon scattering time (about 1 *ps*), and as a result, the *fs* pulses end before the electrons supply its thermal energy to any ions. Heat diffusion is confined to the focal area and, as a result, the precision of the method is increased [84]. In addition, *fs*-laser processing does not require defect electrons for seeding the absorption process; the *fs*-pulse generates enough seed electrons through non-linear ionization [99], making the *fs*-pulsed laser suitable for the precision micromachining applications. Additionally, Zalloum *et al.* [14] showed high quality micromachining with precise geometry can be achieved using *fs*-pulsed laser employing pulse energies that are significantly above diamond's ablation threshold owing to the minimal thermal effects and hydrodynamic expansion of the ablated sample material. Ogawa *et al.* [36] reported that with an increase in energy level, there is an increase in MRR with very minimal change in $sp^2$ graphitised layer thickness.

# 4     DIAMOND SAMPLE PROCESSING

## 4.1    Common fs laser micromachining systems

Ultrashort pulsed laser ablation systems are technologically advanced tools capable of processing various materials at extremely high peak powers (up to few GW), while delivering a highly localized energy that enables a non-thermal ablation of even transparent and low-absorption samples with high precision and accuracy [92]. Active and passive mode-locked oscillators generate ultrashort pulses, that require amplification to reach the desired energy levels to process diamond above its ablation threshold (*i.e.,* at or above 3 J/cm$^2$ [31], see Section 2.1). High peak powers generated in *fs*-pulsed laser system can also inadvertently damage the amplifier, making the *fs*-pulse amplification a challenging task. In passive mode-locking, the energy is transferred from an external source into a gain medium using a semiconductor diode [100], the latter can be of different chemical composition and structure, commonly a synthetic crystal of the garnet group, which enables the generation of discreet wavelengths, pulse durations, output power range and pulse repetition rates in *fs*- pulsed lasers as summarised in Table 1, below.

*Table 1.* Fs-laser systems currently employed in laser micromachining and their characteristics.

| Laser type | Wavelengths (nm) | Pulse durations | Power (W) | Repetition rate (MHz) |
|---|---|---|---|---|
| Ti:sapphire [57, 101-104] | 650 – 1100 | 5 *fs* – 50 *ps* | 0.3 – 1.0 | 80 – 100 |
| Nd:YAG [35, 105] | 1064 | 30 *fs* – 30 *ps* | 0.1 – 1.0 | 10 – 80 |
| Yb:YAG [106-108] | 1030 | 30 *fs* – 30 *ps* | 0.1 – 80.0 [109, 110] | 10 – 80 [111] |
| Yb:KGW [112] | 1064 | 30 *fs* – 30 *ps* | 0.1 – 3.0 [113] | 10 – 80 |
| Fiber [112, 114, 115] | 1030 – 1100 | 50 *fs* – 500 *fs* | 0.05 – 0.1 | 20 – 80 |

Titanium sapphire (Ti:Al$_2$O$_3$) is one of the most common gain medium used in ultra-fast lasers owing to its high gain bandwidth enabling the delivery of the short pulses that are highly scalable to high pulse energy and average power. Neodymium-doped yttrium aluminium garnet (Nd:YAG) and ytterbium-doped (Yb:YAG) gain mediums emit radiation in the N-IR region with Nd$^+$ and Yb$^+$ ions, respectively, providing the lasing activity. The Yb:YAG systems are the highest power systems that use conventional rod, slab and thin disk configuration. Ytterbium-doped potassium gadolinium tungstate (Yb:KGW) compared to Yb:YAG displays large gain bandwidth enabling sub-100 *fs*-pulse durations in mode-locked regime and a low quantum defect that is well below 1.0 %, and a relatively high absorption coefficient, k, of 26 cm$^{-1}$, high emission and absorption cross sections characteristics. Fibre lasers are low power, sub-500 *fs* pulse systems with inherently flexible gain mediums which are normally pumped by semiconductor laser diodes and producing a wavelength in N-IR region. A typical *fs*- laser system is normally equipped with a pump and an oscillator, a chirped pulse amplification (CPA), a ¼- and a ½- wave plates, a beam expander and prism polarizer, a number of steering mirrors, lenses and objectives, a set of light modulators and phase masks, a translational/rotational stage with a sample holder, an light-emitting diode (LED) for illumination of the sample and a charge-coupled device (CCD) camera for viewing and monitoring the sample machining process as schematically shown in Fig. 9.

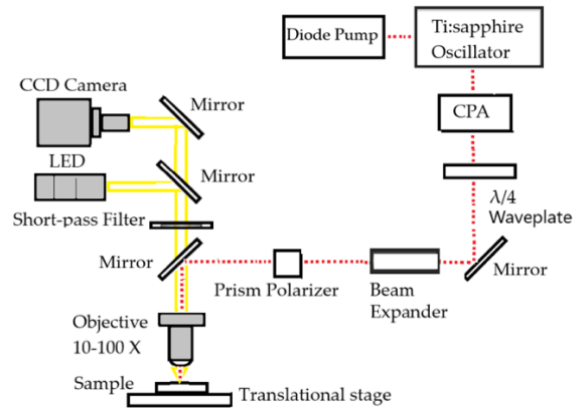

*Fig. 9. Schematic layout of a typical fs-laser system (a Ti:sapphire oscillator is shown).*

Additionally, the *fs-* laser machining system may include a separate sample handling and cleaning stage, and, occasionally, a vacuum chamber equipped with an axillary low pressure pump. Since preliminary $sp^3$ to $sp^2$ phase conversion in diamond and the formation of a graphitised nuclei occurs via a mutually linked process of vaporization and oxidation, laser machining is normally performed in air under abmient conditions, vacuum chambers, however, can be used in instances, where controlled formation of periodic structures on diamond surface is required.

4.2    The effects of fs-laser process parameters on micromachining quality

The laser processing parameters include the following: laser wavelength, $\lambda$ (nm); pulse energy $E_p$ (mJ); pulse duration, $\tau$ (*fs*); pulse repetition rate, $R_p$ (kHz); scanning speed, $V_s$ (mm/s); beam quality, $M^2$; focal length, $f$ (mm); beam diameter, $d_b$ (mm) and numerical aperture of the focusing objective, *NA*. Along with these parameters, some common terms are also used, such as laser peak power, $P_p$ (GW); average power, $P_{avg}$ (mW); beam intensity, $I_p$ (W/cm$^2$); focal spot area, $A_f$ (cm$^2$) and focal spot diameter, $D_f$ (mm).

The peak power $P_p$ is related to $E_p$ and $\tau$ as

$$P_p = \frac{E_p}{\tau} \tag{1}$$

The pulse energy $E_p$ is related to $P_{avg}$ and $R_p$ as

$$E_p = \frac{P_{avg}}{R_p} \tag{2}$$

The beam intensity, $I_p$ which is the energy delivered per unit focal spot area per unit time, is related to $P_p$ and $A_f$ as

$$I_p = \frac{P_p}{A_f} \tag{3}$$

Substituting $P_p$ from Eq. 1 into Eq. 3 gives

$$I_p = \frac{E_p}{\tau . A_f} \tag{4}$$

The fluence, *F,* which is the energy delivered per unit focal spot area, is related to $E_p$ and $A_f$ as

$$F = \frac{E_p}{A_f} \tag{5}$$

The focal spot area $A_f$ is related to $D_f$ as

$$A_f = \frac{\pi D_f^2}{4} \qquad (6)$$

The focal spot diameter $D_f$ is related to $M^2$, $\lambda$, $f$, and $d_b$ as

$$D_f = \frac{4 \cdot M^2 \cdot \lambda \cdot f}{\pi \cdot d_b} \qquad (7)$$

Substituting $D_f$ from Eq. 7 into Eq. 6, gives a relationship between an $A_f$ and $M^2$, $\lambda$, $f$, and $d_b$ as

$$A_f = \pi \cdot \frac{\left(4 \cdot M^2 \lambda \cdot \frac{f}{\pi d_b}\right)^2}{4} \qquad (8)$$

Substituting $A_f$ from Eq. 8 into Eq. 5, gives a governing relationship between the $F$ and $E_p$, $M^2$, $\lambda$, $f$ and $d_b$ as

$$F = \frac{4 \cdot E_p}{\pi \left(4 \cdot M^2 \lambda \cdot \frac{f}{\pi d_b}\right)^2} \qquad (9)$$

The size as well as the intensity distribution profile of the projected beam within the irradiated area affect the material response to the applied laser energy [116]. Also, the laser beam propagation characteristics are known to vary during processing in high power laser delivery systems [117]. A small change in the focal length has a measurable effect on the spot size and the intensity distribution of the projected beam, which in turn influences the stability of the machining process, the local material response and overall the MRR [116]. Also, the size of the laser beam diameter is strongly related to the induced damage probability, overall affecting the damage threshold of the optic system. The smallest diameter of the laser beam permitted for laser induced damage threshold testing as noted in ISO 21254-1:2011 standard is 200 mm [118]. Since higher fluences can be obtained using smaller laser beam diameters many commercial laser machining system suppliers are aiming at providing laser machining centres capable of generating smaller beam diameters. A small micrometre-sized focal spot is achieved by focusing *fs* laser pulses using external lenses, which normally results in a non-linear absorption. For a collimated Gaussian beam focused on a dielectric material, the diffraction-limited minimum waist radius $w_0$, that is ½ of the $D_f$, is given in the Eq. 10, as

$$w_0 = \frac{M^2 \lambda}{\pi (NA)} \qquad (10)$$

Rayleigh range, $z_0$, which is expressed as ½ of the depth of focus, DOF, in a transparent material with retractive index, *n*, and free space wavelength, $\lambda$, is given in Eq. 11, as

$$Z_0 = \frac{M^2 n \lambda}{\pi (NA)^2} \qquad (11)$$

Eqs. 10 – 11 show that a laser beam with a shorter wavelength and a larger *NA* produce a smaller beam waist diameter and a smaller depth of focus, which in turn result in a pulse of a higher fluence.

The number of pulses, *N*, delivered to a specific spot is related to the $D_f$, $R_p$ and $V_s$ as

$$N = \frac{D_f R_p}{V_s} \qquad (12)$$

Eq. 4 shows that a laser beam intensity is directly related to its pulse energy and inversely related to pulse duration and focal spot area. Fluence is linearly related to laser pulse energy and has an inverse square relationship with the beam wavelength (see Eq. 9). A number of pulses delivered at a specific spot on a sample is linearly related to a focal spot diameter and pulse repetition rate, and inversely related to the scanning speed (see Eq. 12). All these parameters influence the characteristics of material ablation and the effect of these laser processing parameters on machining of diamond has been reported in the following experimental studies.

It is known that the laser pulse duration is inversely proportional to its beam intensity( see Eq. 4), and therefore having a strong influence on the machining quality and the MRRs. Ogawa *et al.* [36] compared the surface quality of diamond micromachined by an *fs*- and a *ns*- laser and reported that in an instance when machining was performed using a *ns*- pulsed laser ($\tau$ 140 *ns*, $\lambda$ 1064 nm, $D_f$ 40 mm, $R_p$ 10–100 kHz, $P_{avg}$ <100 W, $E_p$ <1000 mJ), significant amounts of re-solidified materials or debris were left on the machined diamond surface. However, when the surface was micromachined using a *fs*-pulsed laser ($\tau$ 350 *fs*, $R_p$ 200 kHz, $D_f$ 40 mm, $E_p$ <50 mJ, $\lambda$ 1045 nm (IR), $P_{avg}$ < 10 W), no post-machining debris were found as shown in Fig. 10.

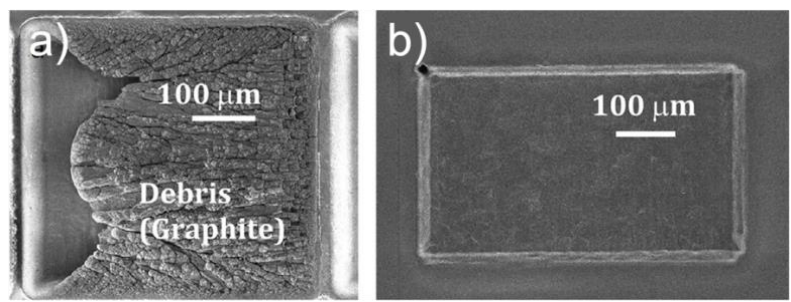

*Fig. 10. Machined surfaces by (a) 140 ns-pulsed and, (b) 350 fs-pulsed laser; adapted from [36].*

It was reported that an *fs*-pulsed laser machining of diamond with circular polarization produces an $R_a$ of ~0.02 μm and an *MRR* of 0.004 mm$^3$/s with no detectable surface layer graphitization as shown in Fig. 11(a) and $R_a$ remains below 0.3 μm range irrespective of the laser power projected on the sample [36]. Notably, the desired high *MRR* and a relatively low $R_a$ during laser machining were only achieved by employing an *fs*- pulsed laser, as shown in Fig. 11(c).

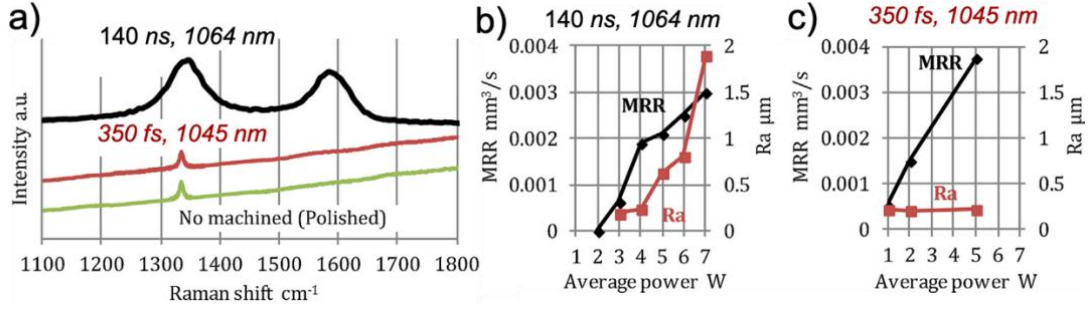

*Fig. 11.* a) Raman spectra after 140 ns- (1064 nm) and 350 fs- (1045 nm) machining; comparative MRR and $R_a$ values for 140 ns- (1064 nm) and 350 fs- (1045 nm) machining operations on dimaond [36].

*Fs*- laser machining with a linear polarized light is normally associated with the formation of so-called light induced periodic surface structures (LIPSS), which appear on the uppermost surface of an ablated workpiece [119]. LIPSSs features vary significantly depending on the selected wavelength and polarization [120, 121], and as shown in Fig. 12(a-c) LIPSS topological features can be spatially reduced from ~0.2 µm (Fig. 12 (a)) to 90 nm (Fig. 12(b)), when an *fs* laser wavelengths is reduced from 800 nm to 400 nm during laser machining of a diamond sample. Additionally, the application of an *fs*- pulsed laser light with a circular polarization results in an $R_a$ reduction from 0.16 µm to 0.096 µm and almost complete disappearance of LIPSS at the same wavelength [36]. The thickness of graphitised $sp^2$-rich layer at the uppermost surface of diamond specimens irradiated using a *ns*-pulsed laser is commonly reported to be a few µm (*e.g.,* 2–5 µm), in contrast, the *fs*-pulsed laser efficiently ablates the material and the thickness of graphitized layer is normally much lower, often only a few tens of nms (*e.g.,* less than 50 nm) as reported by Ogawa *et al.* [36]. Therefore, the *fs*-pulsed laser produces quality ablated surface in diamond with minimal graphitization. A higher pulse energy results in higher fluence, which in turn generates higher *MRRs* (refer to Eq. (9)).

Zalloum *et al.* [14] investigated the effect of varying the pulse energy on the micromachining of HPHT SCD ($\tau$ 200 *fs*, $\lambda$ 800 nm, $R_p$ 250 kHz) and reported no LIPSS-like formation below 75 nJ ablation threshold, which corresponds to a laser fluence of 9.6 J/cm$^2$ using $D_f$ of 1 mm. However, using pulse energies that are significantly above the ablation threshold (*i.e.,* 0.42 mJ), it was shown that *fs* pulses produce clear LIPSS-like saw-tooth structures in diamond at a scale below the diffraction limit (Fig. 12 (c). Increasing the pulse energy to 2.52 mJ increased the depth of LIPSS-like saw-tooth structures (Fig. 12 (d).

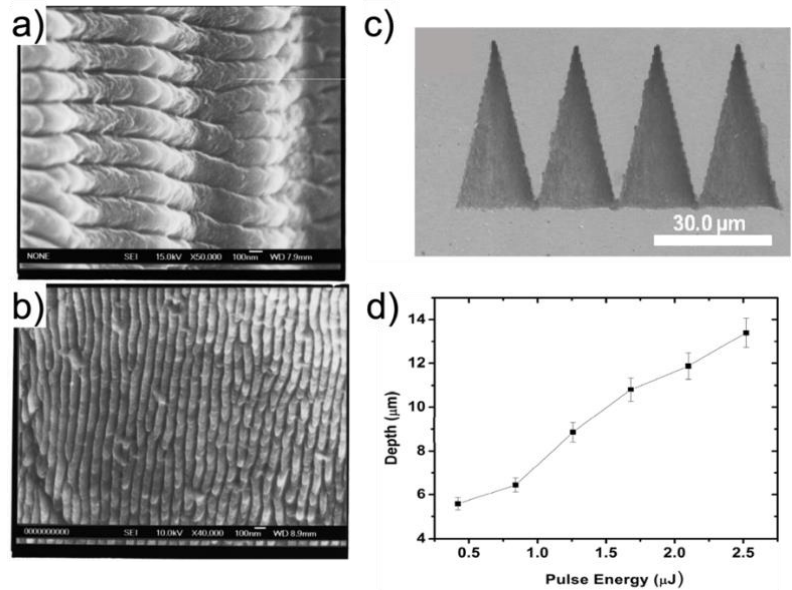

*Fig. 12. a) SEM images of LIPSS on diamond film treated by near normal incidence P-polarized 100 fs- laser with 1 kHz, after 3000 pulses: (a) 800 nm laser with 1.90 J/cm$^2$, (b) 400 nm laser with 0.25 J/cm$^2$, adapted from [36]; (c) formation of LIPPS-like saw-tooth structures during fs-laser micro-structuring, (d) dependence of LIPSS-like saw-tooth depths on fs-pulse energy, adapted from [14].*

By using higher laser beam power and lower scanning speed an improved material ablation is achieved (refer to Eq. (12)). Dou *et al.* [122] showed that the width and depth of an *fs*-laser ($\lambda$ 800 nm, $\tau$ 120 *fs*, $P_{avg}$ 5 W, $R_p$ 1 kHz) induced microgrooves in CVD SCD increased with the increase of laser power and decreased with an increase of scanning speed. The suitable scanning speed for maximizing *MRR* in *fs*-laser machining was suggested to be about 0.1 mm/s [122]. The ablation rate in the depth direction was found to decrease when the scanning speed was greater than 0.3 mm/s [122].

Similarly, the machined structure geometry in *fs*-single shot ablation regime is affected by the fluence as well as an *NA*, which affects the focal spot radius, as given in Eq. (10). For applications that require high depth and high aspect ratio features, it is useful to irradiate the sample sufficiently above its ablation threshold fluence value. The micromachined structures were reported to be spherically symmetric for *NAs* larger than 0.6, for *NAs* lower than 0.6 the resulting structures become asymmetric ($\tau$ 200 *fs*, $\lambda$ 800 nm, $R_p$ 250 kHz) [14].

Ionin *et al.* [103] investigated microscale linear damage tracks inside natural SCD ablated at different *NAs* and focal depths using an *fs*-pulsed laser source ($\lambda$ 744 nm, $\tau$ 120 *fs*, $E_p$ 6.5 mJ, $R_p$ 10 Hz) and reported that there were no detectable bulk damage to the samples at *NA* of 0.17 in 400 – 1600 MW power range, whereas at *NA* of 0.4, 120 *μ*m long continuous graphitic $sp^2$-rich micro-tracks were machined.

Vermeulen *et al.* [102] studied the effects of ionization regime in SCD samples using a range of laser fluences (from 0.29 J/cm$^2$ to 2.35 J/cm$^2$) and power (from 6.92 mW to 55.79 mW) during *fs*-pulsed laser (800 nm, $\tau$ 50 *fs*, $R_p$ 1 kHz, $D_f$ 77.9 mm) ablation and showed that tunnelling ionization was dominant at high (*i.e.,* 2.35 J/cm$^2$) fluences, whereas multi-photon ionization prevailed at low fluences. However, a transition in the ionization regime was observed around the laser fluence of 3.2 J/cm$^2$ from the multi-photon ionization to the tunnelling ionization. The tunnelling ionization regime is believed to enhance the graphitization in the high laser fluence stage where a strong surface cracking phenomenon was also observed. The diameter of the dimple in SCD sample was found to increase from 18.5 mm to 71.9 mm with an increase in the laser power from 10 mW to 260 mW, as shown in Fig. 13. This means that the *MRR* increases with increasing the laser power and incident fluence.

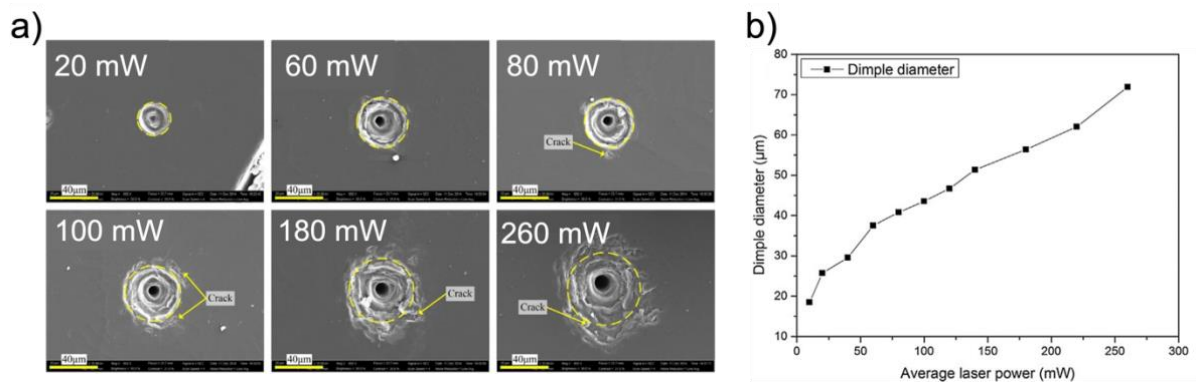

*Fig. 13. (a) Dimples machined with fs-pulsed laser at varying power, (b) A relationship between the dimple diameter and the projected fs-laser power, adapted from [102].*

Likewise, an increase in pulse repetition rate leads to an improved material ablation, due to having a greater number of pulses projected over an irradiated area, for the same pulse energy and pulse duration, as given in Eq. (12).

Sotillo *et al.* [112] studied the effect of repetition rate on the *fs*-laser induced microfabrication on the diamond surface machining double line structures at repetition rates of 5, 25, and 500 kHz, while keeping other parameters constant (*i.e.,* $\tau$ 230 fs, $\lambda$ 515 nm, $E_p$ 200 nJ, $V_s$ 0.5 mm/s, 1.25 *NA*). Their results showed that high repetition rate of 500 kHz resulted in minimal laser-induced graphitization on the sample. The repetition rates not only influence the formation of graphitised $sp^2$-rich layer, but also affect the $R_a$ values of the machined sample. Lower repetition rates induce greater compressive stress in machined samples and facilitate an $sp^2$ fraction formation, which results in a reduced refractive index [123]. White *et al.* [124] also showed higher pulse repetition rates produced drilled holes of larger diameters in PCD diamond films machined using an single and multiple *fs*-laser ($\tau$ 200 fs, $E_p$ 0.8 mJ, *NA* 0.85) pulses, as shown in Fig. 14.

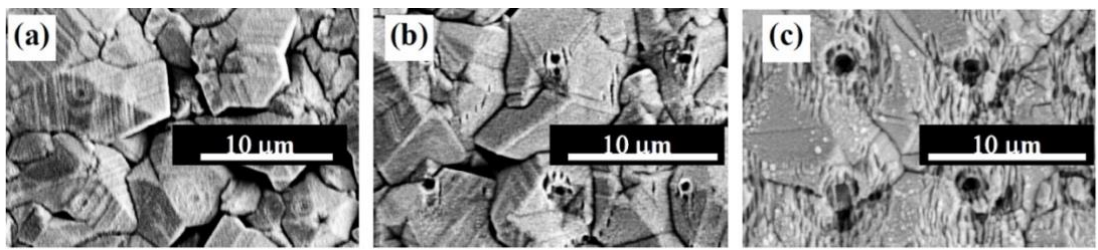

*Fig. 14. Fs-laser structuring of a PCD film surface at with 1 mm beam diameter: (a) single pulse; (b) 50 pulse at 1 kHz and (c) 750 pulses at 250 kHz machining.*

The presence of intrinsic and extrinsic defects and impurities in the synthetic diamond serve as absorption sites that reduce its ablation threshold and need to be considered while comparing different *fs*- pulsed laser ablation results on seemingly similar diamond samples.

Kononenko *et al.* [12] showed that the impurities and defects in the CVD diamond sample contributed to the process of ionization and increased the seed electrons production for the impact ionization, overall reducing the damage threshold for CVD diamond samples by ~30 % compared to a pure, defect-free, natural diamond. The reported damage thresholds for the CVD diamond and the natural diamond sample were 10 – 80 J/cm$^2$ and 2 – 4 J/cm$^2$, respectively for *fs*-laser irradiation (800 nm, 120-*fs*, 8 mm objective; beam waist diameter 3.0 µm).

The laser process parameters, including the wavelength, pulse energy, power, pulse duration and repetition rates, the scanning speed, numerical aperture and beam polarization are to be selected carefully to maximise *MRRs*, attain low $R_a$ values and produce minimal *sp$^2$*-graphitised layer in processed diamond samples. One of the most important laser process parameters is pulse duration, $\tau$, which influences the intensity of the incident beam (refer to Eq. (4)) and, thus, affecting the *MRR*, $R_a$ and *HAZ* mostly. For a specific pulse duration regime, the next most important factor is the laser fluence, *F*, which depends on the $E_p$, $M^2$, $\lambda$, $f$ and $d_b$ (refer to Eq. (9)). In the ultrashort *fs*-pulsed regime, the ablation threshold for diamond is over 3 J/cm$^2$ [31]. The third most important factor is the number of pulses delivered at a specific spot, *N*, which is influenced by $D_f$, $R_p$ and $V_s$ (refer to Eq. (12)). Hence, the pulse duration, fluence and number of pulses at specific spot are the three most important factors dictating the material ablation.

4.3 Post-processing of the *fs* laser irradiated diamond samples

The *fs*-laser machining is a relatively clean process, however a small graphitic debris still may form on the machined surface of diamond, which requires removal.

The most economical and the simplest method is a chemo-mechanical scrub, in which Q-tips, wipes or soft brushes soaked in hydrocarbon solvents such as ethanol, methanol, acetone, etc. and/or demineralised water with anionic cleaning surfactants are employed to manually dislodge and remove the debris [14, 107].

The acid-scrub involves cleaning the samples in a solution of sulphuric ($H_2SO_4$), perchloric and nitric ($HNO_3$) acids (mixed at 5:3:1 wt.%) and heated to 200 °C to remove the graphitized layer formed on the machined surface [35]. Likewise, an aqua regia, a mixture of $HNO_3$ and hydrochloric acids (1:3 wt.%) can be employed to boil the samples at 120 °C for 2 hrs [106]. Following the acid cleaning process, the samples are rinsed in deionised water and dried with compressed nitrogen gas [106].

Ultrasonic cleaning [114, 125] enables removal of debris from machined diamond samples by using cavitation bubbles produced by high frequency (15–400 kHz) sound waves to agitate the liquid media in which the samples are placed for the duration of the US cleaning cycle. Higher frequency produces smaller nodes between the cavitation points resulting in a more precise cleaning of finer features. Often diamond samples are ultrasonically cleaned in $H_2SO_4$ or $HNO_3$ acids for 0.5 – 1 hr, followed by deionised water rinse [114, 125].

One of the advantages of *fs*-laser machining is the easy, economical and uncompromising way by which the samples can be cleaned compared to *ns*- or *ps*-laser machined samples. The latter require more rigorous post-processing methods to remove the fused $sp^2$-debris including the use of multiple repeated steps of chemo-mechanical, acid-scrub and an ultrasonic cleaning and even mechanical grinding [82] and high pressure gas jet ablation [80].

## 4.4  Characterisation of micromachined diamond samples

In industrial settings laser irradiated diamond samples are typically analysed employing a limited range of optical and electron microscopy tools to characterise the 3D surface topography and $R_a$ and spectroscopic techniques to characterise their chemical composition. These techniques vary by their processing capabilities and scalability and are often employed for batch sample screening.

The pattern and geometry of laser machined surfaces is normally visually inspected using an optical microscope equipped with a high magnification (*i.e.,* 100X) objective lenses [35, 82, 104, 106, 107]. The diffraction limits the spatial resolution of examined features to ~200 nm (~½ of blue 400 nm light). A 3D optical microscopy analysis represented by a laser scanning confocal microscopy (LSCM) and a white light interferometry (WLI) enables a detailed observation of surface profile, 3D surface topography and the ablated structures [114] by using a light directed in such a way that it detects the 3D surface. The capabilities of LSCM and WLI vary significantly to batch sample processing, however both the LSCM [115] and the WLI [107, 125] offer a sub-$\mu$m spatial resolution and detailed measurement and analysis of 3D surface topography.

Mechanical surface profilometry uses a contact physical probe with contact force feedback and allows a limited but a straightforward characterisation of surface morphology at ~10 $\mu$m resolution in samples were surface features are highly pronounced and surface roughness is high [126]. To a lesser degree, atomic force microscopy (AFM) analysis with a spatial resolution of a few-nm allows precise surface topography studies including the LIPSS features [106], but at a much reduced sample processing rates and increased costs.

SEM with spatial resolution of sub-50 nm generates the image of a sample by detecting the reflected electrons focusing only on the surface. SEM provides a relatively simple, relatively quick and less costly method to study samples with diverse geometries [107, 125, 127]. However, application of SEM to diamond batch processing is limited as the measurements require a compensation for a non-conducting (*i.e.,* diamond) sample charge accumulation [128].

By using surface profile data the *MRR* can be calculated by (i) multiplying cross-sectional area by the scanning speed [125] and/or, (ii) dividing the total material removal volume by the laser machining time [82, 129, 130].

The physico-chemical and structural characterization of the laser irradiated diamond samples is often performed using Raman and Fourier Transform IR (FT-IR) spectroscopy analysis. The structural transformation in samples can be distinguished by the unique Raman vibrational modes in visible (442 - 633 nm) range at ~1332 cm$^{-1}$ and ~1580 cm$^{-1}$ for tetrahedral diamond and graphite, respectively [35, 36, 104, 114, 131, 132]. Multiwavelength Raman in UV – N-IR range offers significantly enhanced probing capabilities since 244 nm (~ 5 eV) and 325 nm (~3.8 eV) UV Raman directly probes $sp^3$ density of states (DOS) making the signature 'T' resonance mode at ~1030 cm$^{-1}$ [49, 132-134] directly visible. The 785 nm (~1.5 eV) IR Raman allows an enhanced screening for possible polymeric inclusions [133-135] that are often present in CVD NCD samples. Raman measurements allow quantitative and qualitative estimation of the degree of thermal damage to the samples, to monitor the degree of crystallinity/amorphization and with a relative accuracy, to quantitatively estimate the thickness of the $sp^2$ rich layer [136] on laser irradiated samples. Raman in N-IR – visible range is only sensitive to homo-nuclear bonds (*e.g.,* C-C and C=C) of $sp^2$-fraction and therefore its application to precise screening of hydrogenated diamond samples is limited, unless complimented with a UV Raman with its capability to directly probe $sp^3$ DOS. FT-IR, on the other hand, allows the measurement of protonated and hetero-nuclear functional groups (*e.g.,* -OH) and as a traditional and a well-established method FT-IR often compliments Raman when used to classify diamonds [137].

Other analytical techniques such as X-ray photoelectron spectroscopy (XPS) [49, 138, 139], electron spectroscopy for chemical analysis (ESCA), X-ray powder diffraction (XRD) [127] despite offering an enhanced capability to identify $sp^2$ and $sp^3$ phase and surface and subsurface chemical compositions in samples are used sparingly in industrial settings owing to their low batch processing capabilities and a rather involved analytical methods.

5.  APPLICATIONS OF *FS*-LASER MICROMACHINED DIAMOND

Laser micro-patterned diamond tools [140] that are used to mechanically machine multiple micro-patterns on ultra-hard workpiece upon direct contact can be effectively fabricated using *fs*-laser irradiation process [130]. In their production, a high intensity spatial light modulator, such as digital micro-mirror device is employed to project complex patterns into a diamond workpiece sample. This method of production is also making its way to fabrication of identification tags on diamond and direct machining of ultra-hard ceramic materials [141]. *Fs*-pulsed laser micromachining of diamond can potentially be employed for fabrication of diamond pencils or micro grinding wheels to produce MEMS [114], light emitting diodes [142, 143], precision photonic components, flat panel display components [57] including, nano- and micro-scale photonic channels. *Fs*-laser machining has an immense potential to be employed in the fabrication (*i.e.,* drilling) of ultra-precise circular holes in pre-indented diamond anvil cell (DAC) gaskets used in DAC ultra-high-pressure devices [144, 145] and photonics systems such as birefringent regions and Bragg gratings [146, 147]. Diamonds are the prime candidate materials for X-ray compound refractive lenses [148], since they are not only able to withstand extreme radiation and thermal loads but also provide an exceptionally effective and robust optical focusing solution [149, 150]. To manufacture such X-ray lenses, the *fs*-laser micromachining is the only technique devoid of any known technological and manufacturing drawbacks but that is capable of producing lenses with the lowest surface roughness [151].

Nitrogen-vacancy (NV) centres, which occur when a N atom and a vacancy replace two adjacent sites in diamond crystal lattice [152], have shown a potential for quantum computing and sensing applications [150, 153, 154], since NV centres can be optically found, read out and manipulated owing to their long electron-spin coherence times [155-158]. However, the intrinsic inertness of diamond is a significant obstacle to the fabrication of NV-integrated optical systems. The application of *fs*- laser pulses with high repetition rate can solve this problem by machining optical waveguides in diamond and connecting multiple diamond NVs together [112] since there are established routes to fabrication of single mode waveguides the visible to the IR *in situ* in diamond crystal that enable applications such as evanescent field sensors, quantum information systems, and magnetometry [112]. Since *fs*-laser processing produces a reversible but controllable change in the refractive index in diamond membranes, these changes can be effectively exploited for sensing or optical engineering applications [159]. The LIPSS that appear on the surface of diamond open new application avenues such as micro–solar cells devices [160, 161].

In semiconductor manufacturing, diamond can be efficiently sliced into thin semiconductor wafers, a process that is by all means quite challenging, owing to its hard and brittle nature. Since *fs*-laser internal processing converts an inner layer of diamond to an $sp^2$ graphite, which can be easily separated, a fabrication of ultra-thin diamond wafers of 1×1 mm$^2$ is possible [162].

Likewise, by using *fs*-laser irradiation it is possible to fabricate deep-buried graphitic pillars inside the diamond [108] that can function as 3D detectors for hadron therapy to reach deep tumours without damaging the nearby healthy tissues [163]. Conduction in such detectors occurs through highly conductive buried $sp^2$ graphitic channels that can be fabricated with a diameter of ~1.5 *μ*m and ~1150 *μ*m long [12].

Non-thermal photo-ablation of diamond produces negative electron affinity (NEA), an effect by which a conduction band is shifted higher in energy than the vacuum level [164]. Diamond surfaces terminated with H- display NEA only up to ~700 °C (above 700 °C H- is cleaved), making them unsuitable for diamond-based thermionic applications [165], where high temperatures are converted into electricity via an electron emission process. Diamond-based thermionic energy converters that have an emitter (*i.e.,* cathode) made of 'black diamond' [166] absorbing more than 99.9% of the light that strikes them could possibly be fabricated using an *fs*-laser ablation simultaneously with a nano-textured LIPSS surface that is required to harvest solar radiation in diamond-based proton-enhanced thermionic emission (PETE) devices [167]. Evidently, the *fs*-laser-induced properties in diamond are technologically significant and await many other industrial applications.

*Fs*-laser processing technology offers new opportunities for unique applications owing to its clean and essentially non-thermal processing nature which produces minimal *HAZs*, the lowest technologically attainable surface roughness and allows processing of transparent materials through a non-linear absorption mechanism. Compared to the conventional *ns*- and *ps*-laser ablation processes, the *fs*-machining can also generate exceptionally high peak powers and ablate diamond samples with minimal graphitization.

# 6. SUMMARY AND FUTURE RESEARCH PROSPECTS

6.1 The findings of this review can be summarized as follows:

1. *Fs*-laser irradiation can precisely alter the surface topography and transform the chemical structure of diamond depending on the beam characteristics, its intensity and polarization. The application of multi-photon absorption using a focused *fs*-pulsed laser with a high spatial and temporal photon density is able to generate high *MRRs*;

2. The *HAZ* and $R_a$ can be effectively minimised by using an *fs*-pulsed laser owing to its ultra-short interaction time. An *fs*-pulse transfers its energy via a photon-electron coupling prior to phonon-driven thermal diffusion;

3. The laser induced ablation is significantly improved by the presence of intrinsic and extrinsic defects in diamond, which serve as absorption sites;

4. Among all the laser systems, Ti:Sapphire laser can generate the shortest pulses (as short as 5 *fs*) in mode-locked regime of operation and can be exploited in future applications;

5. The laser process parameters are to be selected carefully to maximise *MRRs*, attain low $R_a$ and produce minimal $sp^2$ graphitized region. Among all, the pulse duration, fluence and number of pulses at specific spot are the three (3) most important factors dictating the response of the material to the incident laser energy;

6. The *fs*-laser processing is a relatively clean process generating minimal $sp^2$ graphitized layer and requires minimal post-processing;

7. 3D surface topography and $R_a$ can be precisely studied using 3D optical microscopy measurements. *MRR* can then be calculated from the 3D surface profile. Chemical composition of the micromachined samples can be precisely studied using UV Raman, and especially multiwavelength Raman, spectroscopy;

8. Applications of the *fs*-laser processing are diverse owing to its unique characteristics, such as non-linear absorption and non-thermal nature, that extend its application to processing of transparent and ultra-hard materials.

6.2 Future research prospects

The future research prospects are identified as follows:

1. Laser process parameters need to be optimized to maximize the *MRR*, minimize $R_a$ and further reduce the occurrence of *HAZs* in processed diamond samples;

2. Theoretical framework to accurately describe the response of diamond to the *fs*-laser pulses need to be further refined;

3. *Fs*-laser ablation technique needs to be further developed to be able at 250 nm (~5 eV) wavelength to effectively excite *sp³* DOS.

AUTHOR CONTRIBUTIONS

BAK carried out the search and collection of review data, carried out data analysis, drafted the manuscript. MR designed, conceptualised and coordinated the study, carried out data analysis, drafted and critically revised the manuscript. IVL revised the manuscript. All authors gave the final approval for publication and agree to be held accountable for the work performed therein.

CONFLICT OF INTEREST

The authors confirm that there are no known conflicts of interest associated with this publication and there has been no significant financial support of this work that could have influenced its outcome.